\documentclass[cits]{PoS}\usepackage{amsmath} % HEPHY-PUB 930/13 (2013)
\title{Resolving the Puzzle of the Pion--Photon Transition Form Factor}
\ShortTitle{Resolving the Puzzle of the Pion--Photon Transition
Form Factor}\author{Wolfgang Lucha\\Institute for High Energy
Physics, Austrian Academy of Sciences, Nikolsdorfergasse 18,
A-1050 Vienna, Austria\\E-mail: \email{Wolfgang.Lucha@oeaw.ac.at}}
\author{\speaker{Dmitri Melikhov}\\Institute for High Energy
Physics, Austrian Academy of Sciences, Nikolsdorfergasse 18,
A-1050 Vienna, Austria,\\Faculty of Physics, University of Vienna,
Boltzmanngasse 5, A-1090 Vienna, Austria, and\\D.~V.~Skobeltsyn
Institute of Nuclear Physics, Moscow State University, 119991,
Moscow, Russia\\E-mail: \email{dmitri\_melikhov@gmx.de}}

\abstract{We investigate the form factors for
pseudoscalar-meson--photon transitions by means of dispersive QCD
sum rules and demonstrate that most of the measurements done so
far (in particular, those by {\sc BaBar} for $\eta,$ $\eta'$ and
$\eta_c$ and by Belle for $\pi^0$) are, on the one hand,
compatible with each~other and with the saturation required by
factorization theorems obtained from perturbative QCD and,~on the
other hand, give a hint that saturation is effective already at
relatively low momentum transfers~$Q^2$; this hypothesis is
supported by experimental data for the charged-pion elastic form
factor available in the range
$Q^2\approx2\mbox{--}4\;\mbox{GeV}^2.$ The only exception are the
{\sc BaBar} results for the $\pi^0\gamma$ transition form factor,
which do not fit into such picture. We point out that results
expected from SHMS~at JLab on the pion elastic form factor in the
region $Q^2\approx5\mbox{--}8\;\mbox{GeV}^2$ will provide the
ultimate test~of saturation and factorization and strongly impact
our general view of the form factors up to infinitely
large~$Q^2.$}

\FullConference{The European Physical Society Conference on High
Energy Physics -- EPS-HEP2013\\18--24 July 2013\\Stockholm,
Sweden}

\begin{document}\section{Introduction: Two-Photon Fusion to
Pseudoscalar Mesons}Transitions $\gamma^*\,\gamma^*\to P$ of two
virtual photons $\gamma^*$ into some pseudoscalar meson
$P=\pi^0,\eta,\eta',\eta_c$ are processes that provide important
tests of the dynamics of exclusive processes in QCD. Recently, a
large amount of experimental data on this kind of reaction has
been gathered \cite{cello-cleo,babar,babar2010,babar1,belle}. The
transition amplitude enjoys a particularly simple structure and
involves only a single form factor,
$F_{P\gamma\gamma}(q_1^2,q_2^2)$:
$$\langle\gamma^*(q_1)\,\gamma^*(q_2)|P(p)\rangle={\rm
i}\,\epsilon_{\epsilon_1\epsilon_2q_1q_2}\,
F_{P\gamma\gamma}(q_1^2,q_2^2)\ .$$The behaviour of the form
factor $F_{P\gamma\gamma}(q_1^2,q_2^2)$ at asymptotically large
spacelike momentum transfers $q_1^2\equiv-Q_1^2\le0,$
$q_2^2\equiv-Q_2^2\le0$ results from a QCD factorization theorem
by Lepage and Brodsky~\cite{brodsky}:
$$F_{P\gamma\gamma}(Q_1^2,Q_2^2)\xrightarrow[Q_{1,2}^2\to\infty]{}
12\,e^2\,f_P\int\limits_0^1\frac{{\rm
d}\xi\,\xi\,(1-\xi)}{Q_1^2\,\xi+Q_2^2\,(1-\xi)}\ .$$The
kinematical situation relevant to experiment is characterized by
an almost on-shell ($Q_1^2\approx0$) and an off-shell ($Q^2\equiv
Q_2^2\ge0$) photon: In this case, the pion--photon transition form
factor depends only on the variable $Q^2\equiv Q_2^2$ and will be
labeled by $F_{\pi\gamma}(Q^2).$ For large $Q^2,$ it behaves
asymptotically~like $$Q^2\,F_{\pi\gamma}(Q^2)\xrightarrow
[Q^2\to\infty]{}\sqrt{2}\,f_\pi\ ,\qquad f_\pi=0.130\;\mbox{GeV}\
.$$Upon taking into account meson mixing effects, similar
relations arise also for the $\eta$ and $\eta'$ mesons.

\section{Dispersive QCD Sum Rules for the $\gamma^*\,\gamma^*\to P$
Transition Form Factor}Let us base our QCD sum-rule analysis of
the $\gamma^*\,\gamma^*\to P$ transition form factor on the
amplitude$$\left\langle0\left|j^5_\mu\right|\gamma^*(q_1)\,
\gamma^*(q_2)\right\rangle=e^2\,T_{\mu\alpha\beta}(p|q_1,q_2)\,
\epsilon^\alpha_1\,\epsilon^\beta_2\ ,\qquad p\equiv q_1+q_2\
,$$where $j^5_\mu$ is the axial-vector quark current and
$\epsilon_{1,2}$ are the polarization vectors of the photons. Out
of the four independent Lorentz structures of this amplitude (cf.\
Refs.~\cite{blm2011,lm2011}), here only one~is~relevant:
$$T_{\mu\alpha\beta}(p|q_1,q_2)=p_\mu\,\epsilon_{\alpha\beta
q_1q_2}\,{\rm i}\,F(p^2,Q_1^2,Q_2^2)+\cdots\ .$$The invariant
amplitude $F(p^2,Q_1^2,Q_2^2)$ can be written as dispersion
integral over $p^2$ (at fixed $Q_1^2,Q_2^2$)
$$F(p^2,Q_1^2,Q_2^2)=\frac{1}{\pi}\int\limits_{s_{\rm
th}}^\infty\frac{{\rm d}s}{s-p^2}\,\Delta(s,Q_1^2,Q_2^2)$$of the
physical spectral density $\Delta(s,Q_1^2,Q_2^2),$ with lower
endpoint fixed by the physical threshold $s_{\rm th}.$
Perturbative QCD provides the spectral density as an expansion in
powers of the strong coupling~$\alpha_{\rm s}$:$$\Delta_{\rm
pQCD}(s,Q_1^2,Q_2^2|m)=\Delta^{(0)}_{\rm
pQCD}(s,Q_1^2,Q_2^2|m)+\frac{\alpha_{\rm
s}}{\pi}\,\Delta^{(1)}_{\rm
pQCD}(s,Q_1^2,Q_2^2|m)+\frac{\alpha^2_{\rm
s}}{\pi^2}\,\Delta^{(2)}_{\rm pQCD}(s,Q_1^2,Q_2^2|m)+\cdots$$also
involves the mass $m$ of the quark propagating in the loop spanned
by $j^5_\mu$ and the photon vertices. This 1-loop triangle
diagram, with $j^5_\mu$ and two electromagnetic currents at its
vertices, contributes to $\Delta_{\rm pQCD}(s,Q_1^2,Q_2^2|m)$ the
lowest-order term $\Delta^{(0)}_{\rm pQCD}(s,Q_1^2,Q_2^2|m)$
\cite{1loop}; the 2-loop $O(\alpha_{\rm s})$ term, due~to~the
exchange of a gluon between two quark legs, vanishes \cite{2loop};
the 3-loop $O(\alpha_{\rm s}^2)$ term is non-zero~\cite{3loop}.
\newpage

For low $s$ values, the \emph{physical\/} spectral density differs
from the \emph{perturbative\/} $\Delta_{\rm
pQCD}(s,Q_1^2,Q_2^2|m)$ as the former involves meson pole and
hadron continuum. In the $I=1$ channel, for instance, it reads
$$\Delta(s,Q_1^2,Q_2^2)=\pi\,\delta(s-m_\pi^2)\,\sqrt{2}\,f_\pi\,
F_{\pi\gamma\gamma}(Q^2_1,Q_2^2)+\theta(s-s_{\rm
th})\,\Delta^{(I=1)}_{\rm cont}(s,Q^2_1,Q_2^2)\ .$$

\emph{QCD sum rules\/} enable us to construct relations between
properties of ground-state hadrons and spectral densities of QCD
correlation functions by equating QCD- and hadron-level
representations of $F(p^2,Q_1^2,Q_2^2),$ performing a \emph{Borel
transformation\/} $p^2\to\tau$ to a new ``Borel'' variable $\tau$
to suppress the hadron continuum, and implementing {\em
quark--hadron duality\/} by introducing a low-energy cut on the
spectral representation \cite{lms1,lms2}. Potentially dangerous
nonperturbative power corrections rising with $Q_{1,2}^2$ are
absent in the local-duality (LD) limit $\tau=0$ \cite{ld}. The
procedure gives for~$F_{P\gamma\gamma}(Q_1^2,Q_2^2)$ $$\pi\,f_P\,
F_{P\gamma\gamma}(Q_1^2,Q_2^2)=\int\limits_{4m^2}^{s_{\rm
eff}(Q_1^2,Q_2^2)}\hspace{-2.7ex}{\rm d}s\,\Delta_{\rm
pQCD}(s,Q_1^2,Q_2^2|m)\ ,$$with all details of nonperturbative-QCD
dynamics encoded in some effective threshold $s_{\rm
eff}(Q_1^2,Q_2^2).$ The actual challenge in this game is to
formulate reliable criteria for fixing effective thresholds
\cite{lms1}.

For notational simplicity, from now on we switch to variables
$Q^2$ and $\beta$ defined by $Q^2\equiv Q_2^2$ and
$0\le\beta\equiv Q_1^2/Q_2^2 \le 1$. For given $\beta,$ our
effective threshold $s_{\rm eff}(Q^2,\beta)$ in the limit
$Q^2\to\infty$ is found by matching to the asymptotic
factorization result. In the general case $m\ne0,$ this gives
$s_{\rm eff}(Q^2\to\infty,\beta)$ as a function of $\beta.$ Only
for massless fermions, $m=0,$ the asymptotic value is reproduced
for all $\beta$ if $s_{\rm eff}(Q^2,\beta)$ behaves like $s_{\rm
eff}(Q^2\to\infty,\beta)=4\,\pi^2\,f_\pi^2.$ The LD model
\emph{assumes\/} that also for finite values of $Q^2,$ $s_{\rm
eff}(Q^2,\beta)$ is tolerably approximated by the $Q^2\to\infty$
limit, i.e., $s_{\rm eff}(Q^2,\beta)=s_{\rm
eff}(Q^2\to\infty,\beta).$ Employing the abbreviation
$F_{P\gamma}(Q^2)\equiv F_{P\gamma\gamma}(0,Q^2)$ for the
``empirical'' pseudoscalar-meson--photon transition form factor,
its LD expression for $Q_1^2=0$ and $m=0$ is given, in the
single-flavour case,~by\begin{equation}\label{srfp}
F_{P\gamma}(Q^2)=\frac{1}{2\,\pi^2\,f_P}\,\frac{s_{\rm
eff}(Q^2)}{s_{\rm eff}(Q^2)+Q^2}\ .\end{equation}Irrespective of
the behaviour of $s_{\rm eff}(Q^2)$ for $Q^2\to0,$
$F_{P\gamma}(Q^2=0)$ is related to the axial
anomaly~\cite{blm2011}.

\section{Local-Duality Approach to Pseudoscalar-Meson--Photon
Transition Form Factors}Let us discuss the various mesons in the
order of increasing amount of troubles to be overcome.

\subsection{Form Factor for the $\eta_c$-Meson--Photon Transition
$\gamma^*\,\gamma^*\to\eta_c$}
\begin{figure}[ht]\begin{center}\begin{tabular}{cc}
\includegraphics[scale=.46897]{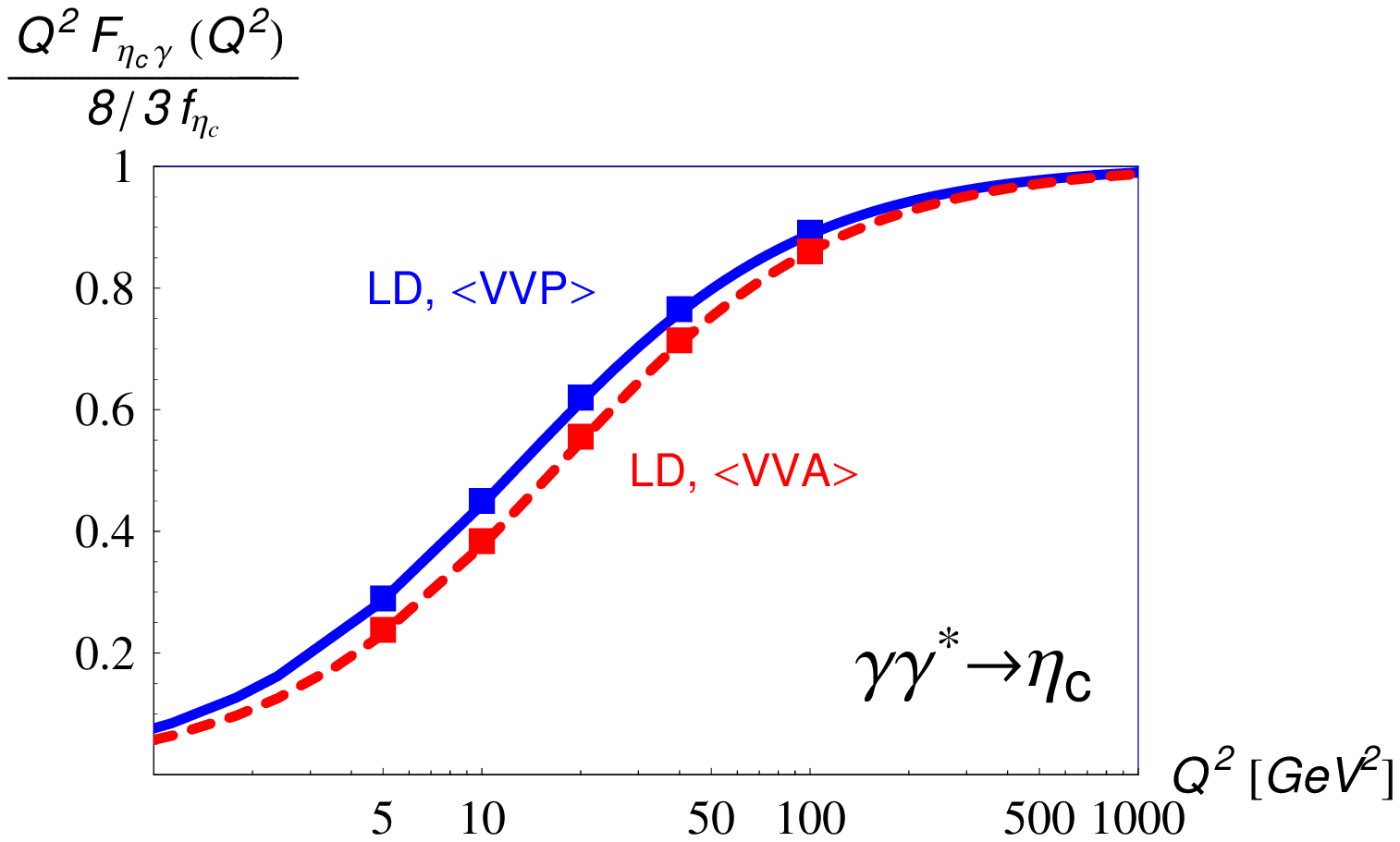}&
\includegraphics[scale=.46897]{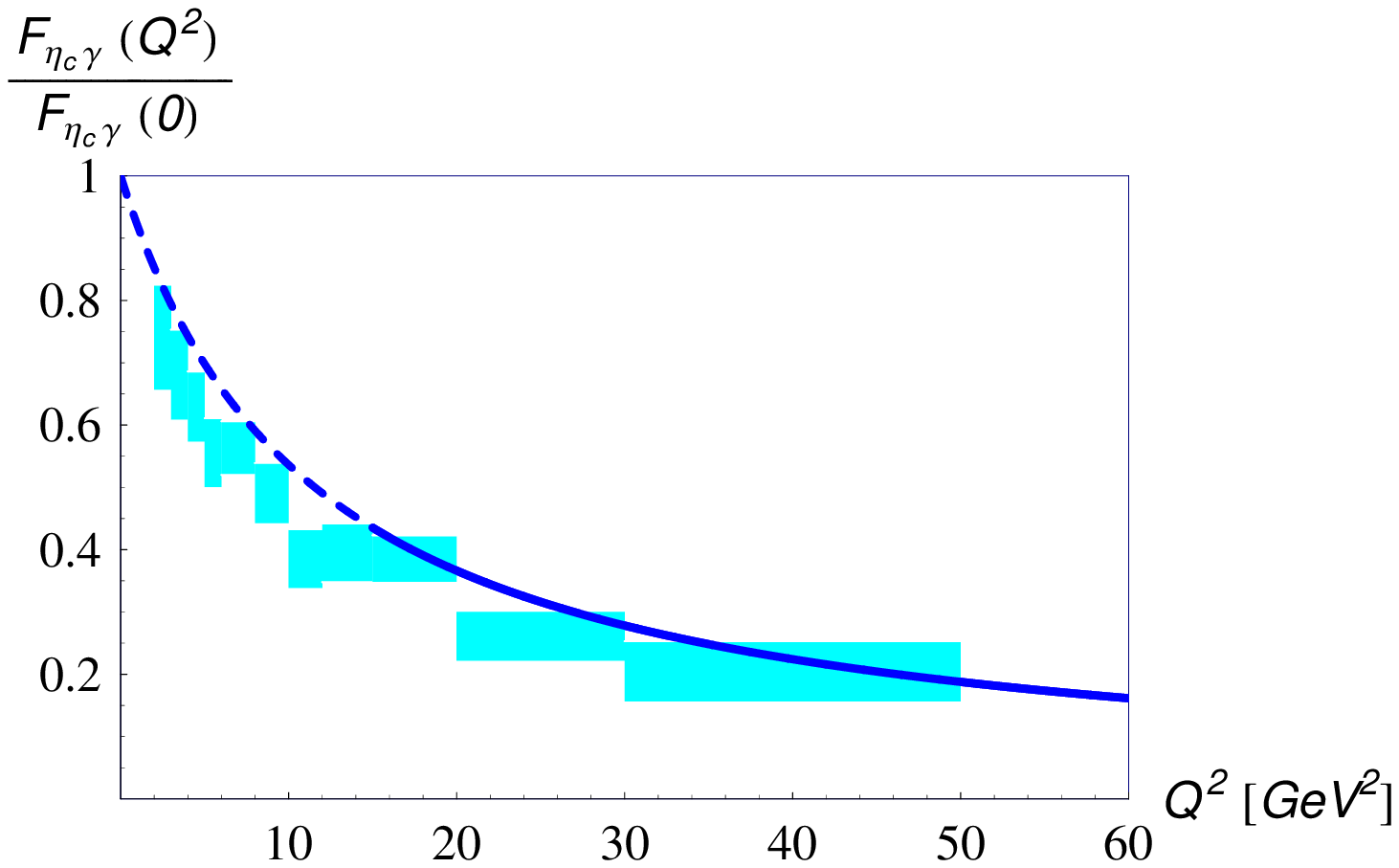}\\{\small
(a)}&{\small (b)}\end{tabular}\caption{Form factor for the
transition $\gamma\,\gamma^*\to\eta_c$: (a) form factor for finite
$Q^2,$ arising from LD sum~rules for $\langle AVV\rangle$ and
$\langle PVV\rangle$; (b) LD model for the correlator $\langle
PVV\rangle$ confronted with experimental data by {\sc
BaBar}~\cite{babar2010}.}\label{Fig:2}\end{center}\end{figure}

For massive quarks, we may study the
correlators $\langle AVV\rangle$ and $\langle PVV\rangle$ of two
vector currents and either an axial-vector or a pseudoscalar quark
current, respectively, in order to find LD sum rules for the
transition form factor under consideration \cite{lm2011}. By
satisfying the perturbative-QCD~factorization theorem, the
asymptotic values $s_{\rm eff}(Q^2\to\infty,\beta)$ for both
correlators are derived. The exact effective thresholds valid for
$\langle AVV\rangle$ and $\langle PVV\rangle$ prove to differ from
each other \cite{lm2011}. Under the LD hypothesis we obtain the
results depicted in Fig.~\ref{Fig:2}. Clearly, for very small
$Q^2$ the LD model is, by construction, not applicable. Adopting
it, nevertheless, the full way down to $Q^2=0$ gives
$F_{\eta_c\gamma}(0)=0.067\;\mbox{GeV}^{-1}$ from $\langle
AVV\rangle$ and $F_{\eta_c\gamma}(0)=0.086\;\mbox{GeV}^{-1}$ from
$\langle PVV\rangle.$ These outcomes must be confronted with the
corresponding experimental data $F_{\eta_c\gamma}(0)=(0.08\pm0.01)
\;\mbox{GeV}^{-1}$: For both correlators used,~the LD model
performs reasonably well over a broad $Q^2$ range starting at
surprisingly low $Q^2;$ cf.~Ref.~\cite{kroll}.

\subsection{Form Factors for the $\eta^{(\prime)}$-Meson--Photon
Transitions $\gamma\,\gamma^*\to(\eta,\eta^{\prime})$}
\begin{figure}[h]\begin{center}\begin{tabular}{cc}
\includegraphics[scale=.48064]{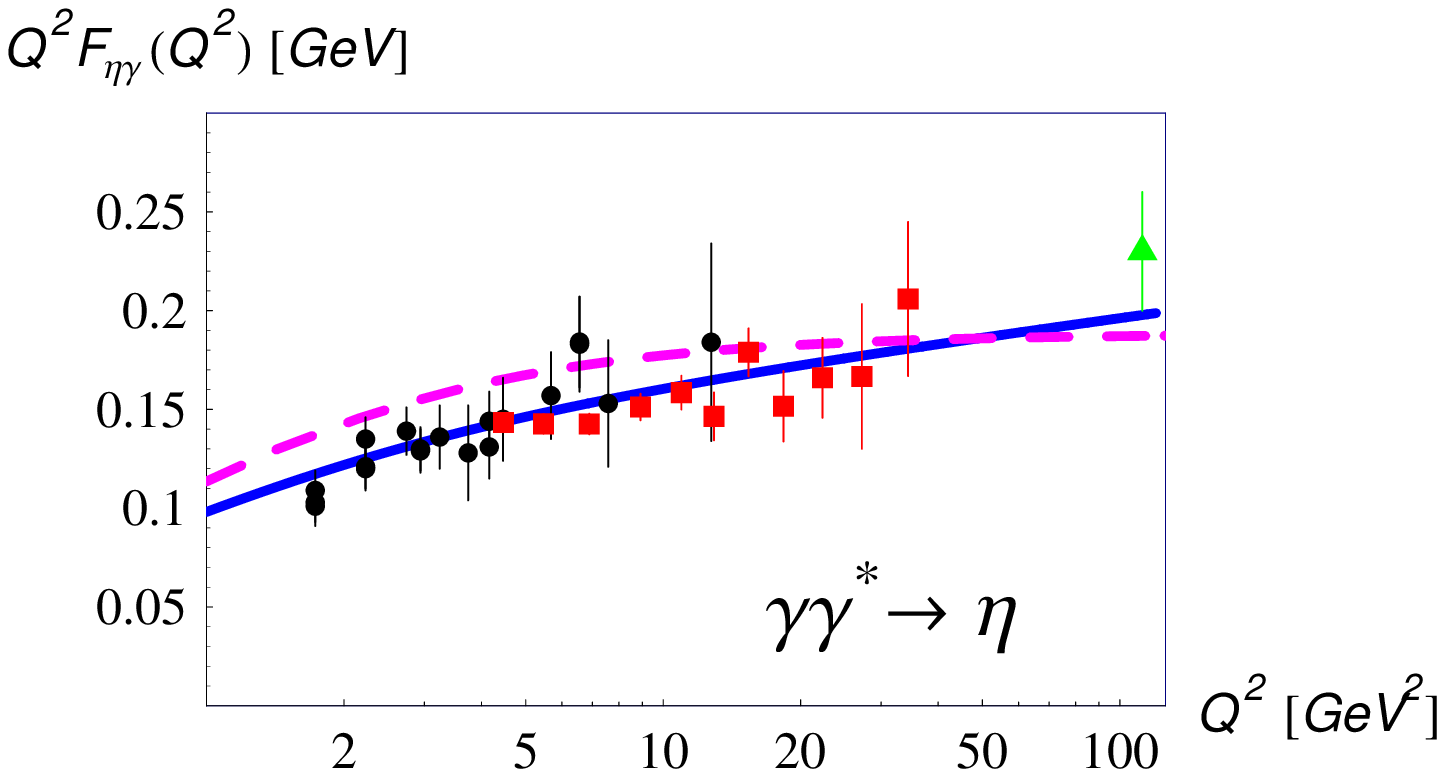}&
\includegraphics[scale=.48064]{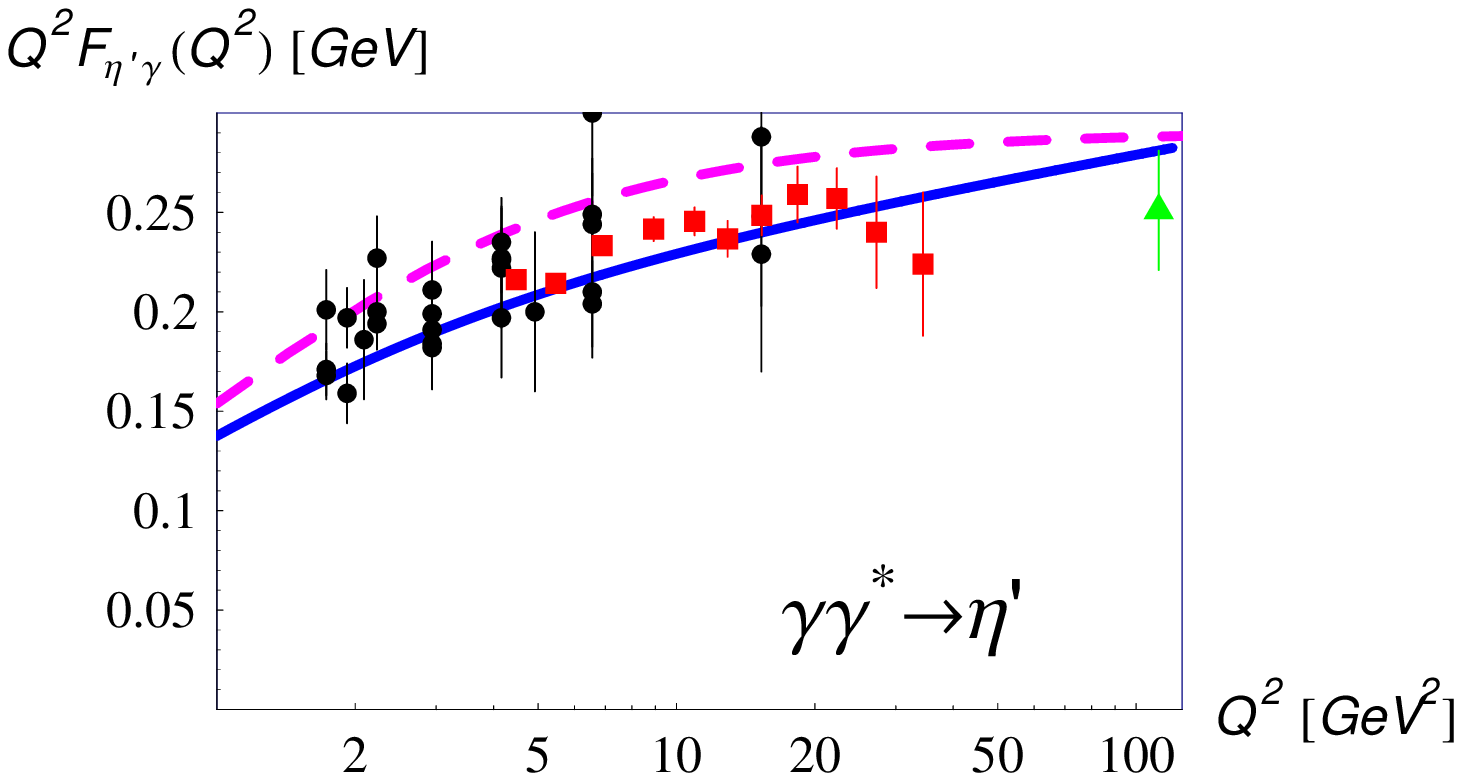}\\{\small
(a)}&{\small (b)}\end{tabular}\caption{Form factors
$F_{(\eta,\eta^\prime)\gamma}(Q^2)$ for the transitions
$\gamma\,\gamma^*\to(\eta,\eta^\prime)$: LD predictions
\cite{blm2011,lm2011} (dashed magenta lines) and recent fits
\cite{ms2012} (solid blue lines) to the experimental data
\cite{cello-cleo,babar1} for $\gamma\,\gamma^*\to\eta$ (a) and
$\gamma\,\gamma^*\to\eta^\prime$ (b).}\label{Fig:3}\end{center}
\end{figure}

A minor
complication arises in any description of two-photon fusion to
$\eta$ and $\eta^\prime;$ the mixing of their nonstrange
$n\sim(\bar uu +\bar dd)/\sqrt{2}$ and strange $s\sim\bar ss$
components has to be taken into account~\cite{mixing}:
$$F_{\eta\gamma}(Q^2)=
F_{n\gamma}(Q^2)\cos\phi-F_{s\gamma}(Q^2)\sin\phi\ ,\qquad
F_{\eta'\gamma}(Q^2)=
F_{n\gamma}(Q^2)\sin\phi+F_{s\gamma}(Q^2)\cos\phi\ ,$$with mixing
angle $\phi\approx38^\circ.$ The LD predictions for the
\emph{flavour\/} form factors $F_{n\gamma}(Q^2)$ and
$F_{s\gamma}(Q^2)$~are$$f_n\,F_{n\gamma}(Q^2)=
\int\limits_0^{s_{\rm eff}^{(n)}(Q^2)}\hspace{-1.7ex}{\rm
d}s\,\Delta_n(s,Q^2)\ ,\qquad f_s\,F_{s\gamma}(Q^2)=
\int\limits_0^{s_{\rm eff}^{(s)}(Q^2)}\hspace{-1.6ex}{\rm
d}s\,\Delta_s(s,Q^2)\ .$$The two separate effective thresholds
$s_{\rm eff}^{(n)}(Q^2)$ and $s_{\rm eff}^{(s)}(Q^2)$ need not be
identical: $s_{\rm eff}^{(n)}(Q^2)=4\,\pi^2\,f_n^2$ and $s_{\rm
eff}^{(s)}(Q^2)=4\,\pi^2\,f_s^2,$ from the LD point of view, with
$f_n\approx1.07\,f_\pi$ and $f_s\approx1.36\,f_\pi$. The outcomes
of this LD model \cite{blm2011,lm2011} show satisfactory agreement
with available experimental data \cite{cello-cleo,babar1}
(Fig.~\ref{Fig:3}).

\subsection{Form Factor for the Neutral-Pion--Photon Transition
$\gamma\,\gamma^*\to\pi^0$}
\begin{figure}[h]\begin{center}\begin{tabular}{cc}
\includegraphics[scale=.518806]{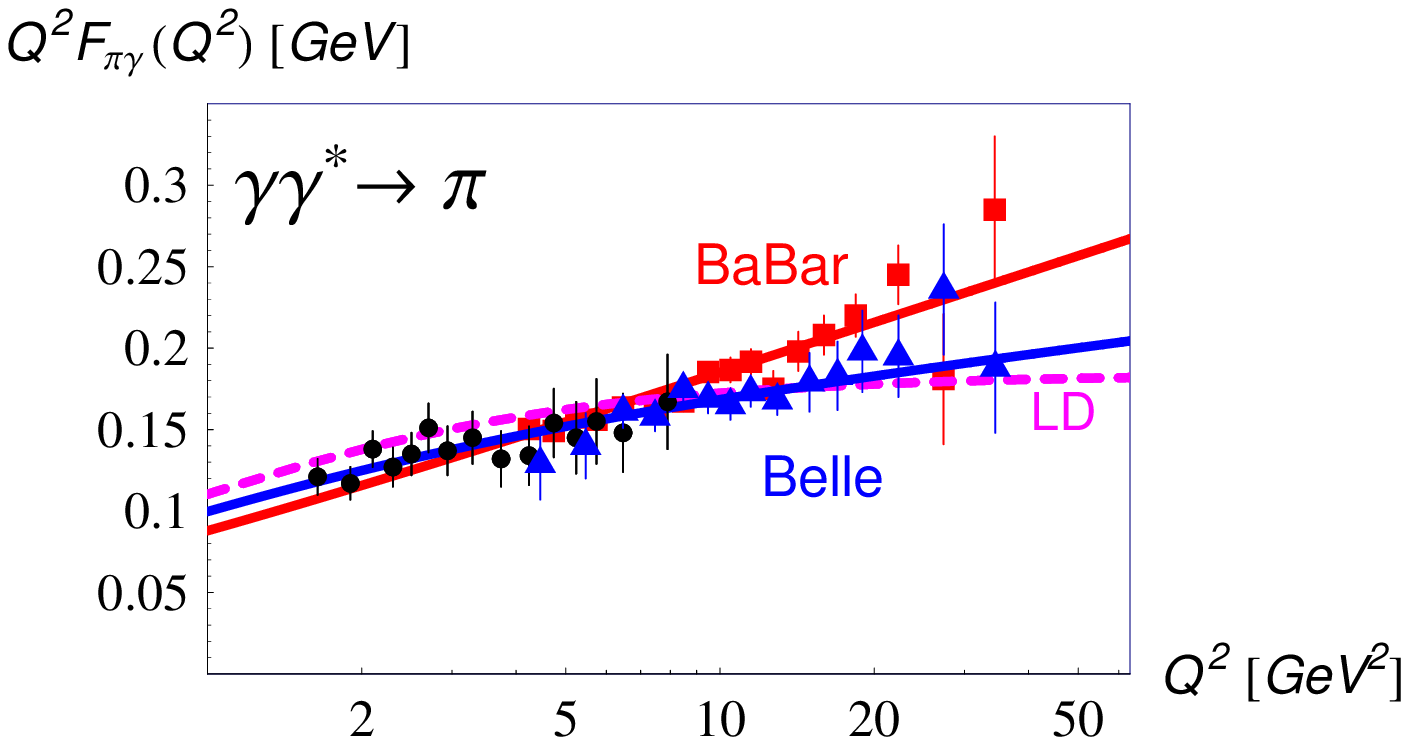}&
\includegraphics[scale=.518806]{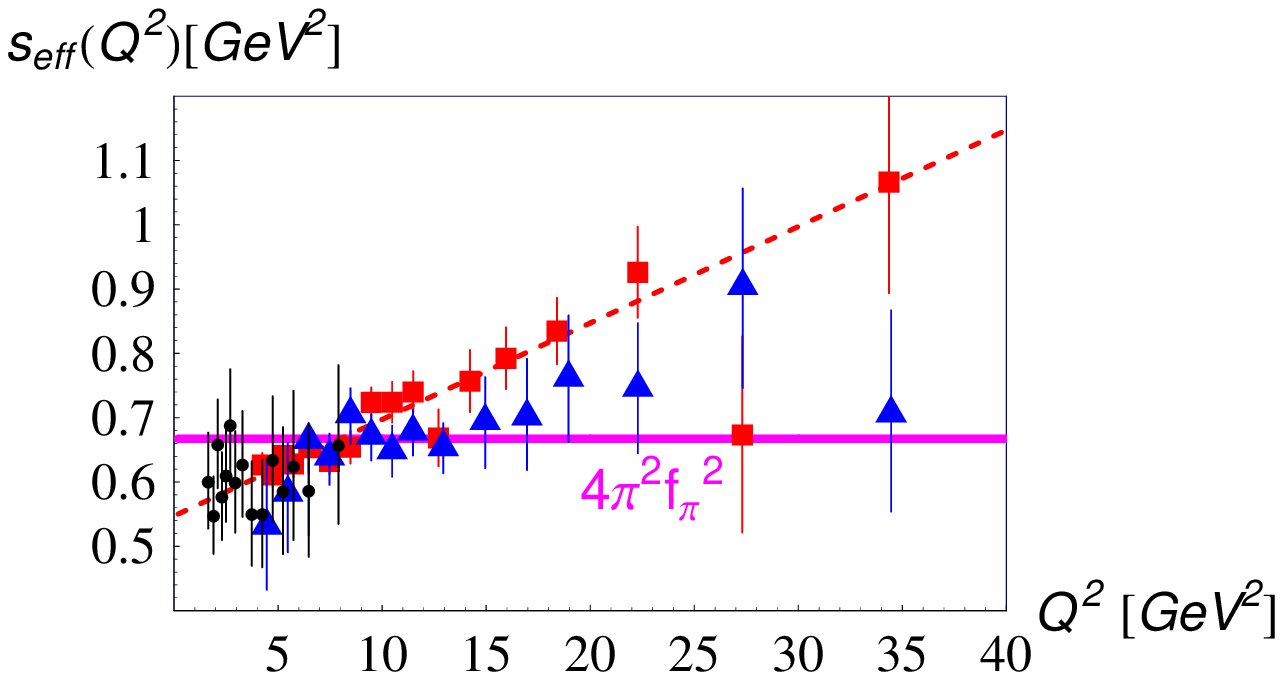}\\{\small
(a)}&{\small (b)}\end{tabular}\caption{LD approach (magenta lines)
to the transition $\gamma\,\gamma^*\to\pi^0$: (a) form factor
$F_{\pi\gamma}(Q^2)$ vs.~fits \cite{ms2012}~(solid lines) to
experiment \cite{cello-cleo,babar,belle}; (b) equivalent effective
threshold $s_{\rm eff}(Q^2),$ fixed for each data point by
Eq.~(\protect\ref{srfp}).}\label{Fig:4}\end{center}\end{figure}

The current situation with the
two-photon fusion to the neutral pion $\pi^0$ constitutes a true
source of worry, even on rather general grounds: The behaviour of
each of the $\eta,$ $\eta'$, and $\pi^0$ transition form factors
for large $Q^2$ is described by spectral densities computed from
perturbative-QCD diagrams; it therefore should be identical for
all light pseudoscalar mesons \cite{ms2012}. This fact becomes
evident~upon noting that the $\langle AVV\rangle$ QCD sum rule in
its LD limit $\tau=0$ is equivalent to the anomaly
sum~rule~\cite{teryaev2}$$2\,\sqrt{2}\,\pi^2\,f_\pi\,
F_{\pi\gamma}(Q^2)=1-2\,\pi\int\limits_{s_{\rm th}}^\infty{\rm
d}s\,\Delta^{(I=1)}_{\rm cont}(s,Q^2)\ .$$The same relations hold,
\emph{mutatis mutandis\/}, also for the $I=0$ and $\bar ss$
channels. Thus, the form factors $F_{\pi\gamma}(Q^2),$
$F_{\eta\gamma}(Q^2),$ and $F_{\eta'\gamma}(Q^2)$ at large $Q^2$
are controlled by the behaviour of the respective hadron continuum
contributions $\Delta_{\rm cont}(s,Q^2)$ for large $s$
\cite{ms2012}. Quark--hadron duality assures us that the latter
must equal their QCD-level counterparts $\Delta_{\rm
pQCD}(s,Q^2),$ which, as purely perturbative quantities, are
identical for all channels. Surprisingly, {\sc BaBar} reports in
the case of the pion transition form~factor a distinct
disagreement both with the $\eta$ and $\eta'$ form factors and
with the LD predictions for $Q^2$ up to
$Q^2\approx40\;\mbox{GeV}^2.$ Still worse, in contrast to our
quantum-mechanical expectations the deviations~from LD predictions
rise with $Q^2$ even in the region $Q^2\approx40\;\mbox{GeV}^2.$
So, QCD has a hard time when trying to understand the {\sc BaBar}
$\pi^0$ results (cf.~also Ref.~\cite{findings}). In this context,
a recent Belle measurement brought a great relief to theory since,
although statistically~consistent with the {\sc BaBar} data
\cite{agaev,pere}, the Belle findings for the $\pi^0\,\gamma$
transition form factor are fully compatible with existing $\eta$
and $\eta'$ data as well as with the very likely onset of the LD
regime already in the range
$Q^2\gtrapprox5\mbox{--}10\;\mbox{GeV}^2$~(Fig.~\ref{Fig:4}).

\section{Elastic Form Factor of the Charged Pion}
\begin{figure}[h]\begin{center}
\includegraphics[scale=.64185]{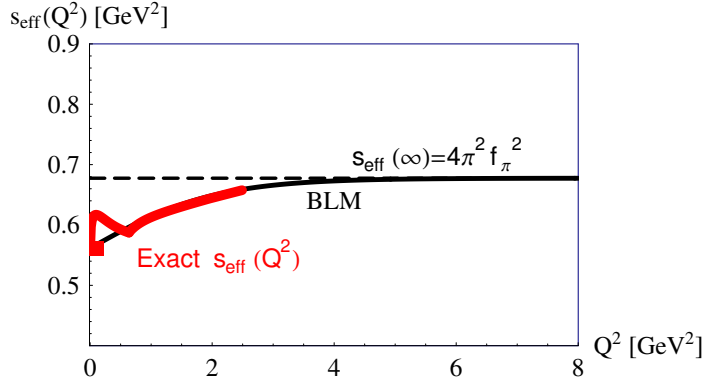}\end{center}
\caption{Effective threshold $s_{\rm eff}(Q^2)$ for the
charged-pion elastic form factor: exact behaviour as function~of
momentum transfer derived from experiment (red line) vs.\ a simple
parametrization labeled BLM by Braguta \emph{et al.} (solid black
line) \cite{blm2011} that approaches its expected asymptotic limit
$s_{\rm eff}(\infty)=4\,\pi^2\,f_\pi^2$ (dashed black line).}
\label{Fig:5}\end{figure}

Our recent
detailed reanalysis of the charged pion's elastic form factor
within the framework of QCD sum rules in LD limit \cite{blm2011}
lends strong support to this LD concept: the exact effective
threshold for this case may be computed from precision
measurements at low $Q^2$ (Fig.~\ref{Fig:5}). Taking~into account
\begin{itemize}\item the experimental results for the effective
threshold $s_{\rm eff}(Q^2)$ at low momentum transfer
$Q^2$~and\item its general feature of converging to the asymptotic
value $4\,\pi^2\,f_\pi^2$ in the region
$Q^2\gtrapprox4\mbox{--}6\;\mbox{GeV}^2,$\end{itemize}there are
good reasons to expect that $s_{\rm eff}(Q^2)$ for the pion's
elastic form factor reaches its asymptotic value already somewhere
near $Q^2\approx4\mbox{--}6\;\mbox{GeV}^2.$ If so, the LD approach
predicts the pion elastic form factor accurately for all
$Q^2\gtrapprox4\mbox{--}6\;\mbox{GeV}^2;$ forthcoming measurements
by JLab's SHMS will tell \cite{Fpi12}.

\section{Summary of Main Results and Conclusions}Our improved QCD
sum-rule approach promotes the effective continuum threshold
entering in the QCD-level spectral representation of the
correlator under consideration to a truly key quantity of the
modified formalism. Determining this object by matching the
predictions of our technique~to the consequences of QCD
factorization theorems, we studied the transition form factors for
two-photon fusion to the $\pi^0,$ $\eta,$ $\eta',$ and $\eta_c$
mesons by QCD sum rules in LD limit, and arrived
at~crucial~insights.\begin{itemize}\item For momentum transfer
$Q^2$ larger than a few $\mbox{GeV}^2,$ this LD model is expected
to reproduce to a satisfactory degree the transition form factors
for all $P\to\gamma\,\gamma^*$ processes: for
$P=\eta,\eta',\eta_c,$ it indeed does; for $P=\pi^0,$ {\sc BaBar}
\cite{babar} requires an LD-violating linear rise of $s_{\rm
eff}(Q^2)$ instead of its approach to a constant
[Fig.~\ref{Fig:4}(b)] but, by confirming the LD claim, Belle
\cite{belle} saves~our~day.\item Assuming applicability of the LD
ideas also to the \emph{elastic\/} form factor of the charged
pion, our findings lead us to suspect that in this case the
accuracy of the LD model increases with $Q^2$ in the region
$Q^2\approx4\mbox{--}8\;\mbox{GeV}^2$ \cite{blm2011}. Accurate
experimental data on the $\pi^\pm$ form factor suggest that the LD
value $s_{\rm eff}(\infty)=4\,\pi^2\,f_\pi^2$ of the effective
threshold is reached already at
$Q^2\approx4\mbox{--}8\;\mbox{GeV}^2;$ this claim looks forward to
its confrontation with the results by the JLab~upgrade SHMS
\cite{Fpi12}.\end{itemize}

\vspace{4ex}\noindent{\bf Acknowledgments.} D.M.\ thanks B.~Stech,
S.~Brodsky, A.~Oganesian and O.~Teryaev for numerous helpful
discussions. D.M.\ was supported by the Austrian Science Fund
(FWF), Project No.~P22843.


\begin{thebibliography}{99}
\bibitem{cello-cleo}H.~J.~Behrend \emph{et al.\/}, Z.~Phys.~C
\textbf{49} (1991) 401; J.~Gronberg \emph{et al.\/}, Phys.~Rev.~D
\textbf{57} (1998) 33.
\bibitem{babar}B.~Aubert \emph{et al.\/}, Phys.~Rev.~D \textbf{80}
(2009) 052002.
\bibitem{babar2010}J.~P.~Lees \emph{et al.\/}, Phys.~Rev.~D
\textbf{81} (2010) 052010.
\bibitem{babar1}P.~del Amo Sanchez \emph{et al.\/}, Phys.~Rev.~D
\textbf{84} (2011) 052001.
\bibitem{belle}S.~Uehara \emph{et al.\/}, Phys.~Rev.~D \textbf{86}
(2012) 092007.
\bibitem{brodsky}G.~P.~Lepage and S.~J.~Brodsky, Phys.~Rev.~D
\textbf{22} (1980) 2157.
\bibitem{blm2011}V.~Braguta, W.~Lucha, and D.~Melikhov,
Phys.~Lett.~B \textbf{661} (2008) 354; I.~Balakireva, W.~Lucha,
and D.~Melikhov, J.~Phys.~G \textbf{39} (2012) 055007;
Phys.~Rev.~D \textbf{85} (2012) 036006;
Phys.~Atom.~Nucl.~\textbf{76} (2013) 326.
\bibitem{lm2011}W.~Lucha and D.~Melikhov, J.~Phys.~G \textbf{39}
(2012) 045003; Phys.~Rev.~D \textbf{86} (2012) 016001.
\bibitem{1loop}J.~Ho\v rej\v s\'i and O.~V.~Teryaev, Z.~Phys.~C
\textbf{65} (1995) 691; D.~Melikhov and B.~Stech,
Phys.~Rev.~Lett.~\textbf{88} (2002) 151601; D.~Melikhov,
Eur.~Phys.~J.~direct \textbf{C4} (2002) 2, arXiv:hep-ph/0110087.
\bibitem{2loop}F.~Jegerlehner and O.~V.~Tarasov,
Phys.~Lett.~B \textbf{639} (2006) 299; R.~S.~Pasechnik and
O.~V.~Teryaev, Phys.~Rev.~D \textbf{73} (2006) 034017.
\bibitem{3loop}J.~Mondejar and K.~Melnikov, Phys.~Lett.~B
\textbf{718} (2013) 1364.
\bibitem{lms1}W.~Lucha, D.~Melikhov, and S.~Simula, Phys.~Rev.~D
\textbf{76} (2007) 036002; Phys.~Lett.~B \textbf{657} (2007) 148;
Phys.~Atom.~Nucl.~\textbf{71} (2008) 1461; Phys.~Lett.~B
\textbf{671} (2009) 445; D.~Melikhov, Phys.~Lett.~B \textbf{671}
(2009) 450.
\bibitem{lms2}W.~Lucha, D.~Melikhov, and S.~Simula,
Phys.~Rev.~D \textbf{79} (2009) 096011; J.~Phys.~G \textbf{37}
(2010) 035003; Phys.~Lett.~B \textbf{687} (2010) 48;
Phys.~Atom.~Nucl.~\textbf{73} (2010) 1770; J.~Phys.~G \textbf{38}
(2011) 105002; Phys.\ Lett.~B \textbf{701} (2011) 82; Phys.~Rev.~D
\textbf{88} (2013) 056011; W.~Lucha, D.~Melikhov, H.~Sazdjian, and
S.\ Simula, Phys.~Rev.~D \textbf{80} (2009) 114028.
\bibitem{ld}V.~A.~Nesterenko and A.~V.~Radyushkin, Phys.~Lett.~B
\textbf{115} (1982) 410.
\bibitem{kroll}P.~Kroll, Eur.~Phys.~J.~C \textbf{71} (2011) 1623.
\bibitem{mixing}V.~V.~Anisovich, D.~I.~Melikhov, and V.~A.~Nikonov,
Phys.~Rev.~D \textbf{55} (1997) 2918; V.~V.~Anisovich, D.\
V.~Bugg, D.~I.~Melikhov, and V.~A.~Nikonov, Phys.~Lett.~B
\textbf{404} (1997) 166; T.~Feldmann, P.~Kroll, and B.~Stech,
Phys.~Rev.~D \textbf{58} (1998) 114006; Phys.~Lett.~B \textbf{449}
(1999) 339.
\bibitem{ms2012}D.~Melikhov and B.~Stech, Phys.~Rev.~D \textbf{85}
(2012) 051901(R); Phys.~Lett.~B \textbf{718} (2012) 488.
\bibitem{teryaev2}Y.~N.~Klopot, A.~G.~Oganesian, and O.~V.~Teryaev,
Phys.~Lett.~B \textbf{695} (2011) 130; Phys.~Rev.~D \textbf{84}
(2011) 051901(R); JETP~Lett.~\textbf{94} (2011) 729; Phys.~Rev.~D
\textbf{87} (2013) 036013; \textbf{88} (2013) 059902(E).
\bibitem{findings}H.~L.~L.~Roberts \emph{et al.\/}, Phys.~Rev.~C
\textbf{82} (2010) 065202; S.~J.~Brodsky, F.-G.~Cao, and G.~F.~de
T\'eramond, Phys.~Rev.~D \textbf{84} (2011) 033001; \textbf{84}
(2011) 075012; A.~P.~Bakulev, S.~V.~Mikhailov, A.~V.\ Pimikov, and
N.~G.~Stefanis, Phys.~Rev.~D \textbf{84} (2011) 034014;
\textbf{86} (2012) 031501(R); \textbf{87} (2013) 094025; G.~F.~de
T\'eramond and S.~J.~Brodsky, arXiv:1203.4025 [hep-ph].
\bibitem{agaev}S.~S.~Agaev, V.~M.~Braun, N.~Offen, and
F.~A.~Porkert, Phys.~Rev.~D \textbf{83} (2011) 054020; \textbf{86}
(2012) 077504.
\bibitem{pere}P.~Masjuan, Phys.~Rev.~D \textbf{86} (2012) 094021;
R.~Escribano, P.~Masjuan, and P.~Sanchez-Puertas, arXiv:1307.2061
[hep-ph].
\bibitem{Fpi12}G.~M.~Huber \emph{et al.\/}, Jefferson Lab PAC 30
Proposal PR12-06-101 ``Measurement of the Charged Pion Form Factor
to High $Q^2$'' (2006).\newpage\end{thebibliography}
\end{document}